# Pervasive Service Architecture for a Digital Business Ecosystem

Thomas Heistracher[1], Thomas Kurz[1], Claudius Masuch[1], Pierfranco Ferronato[2], Miguel Vidal[3], Angelo Corallo[4], Gerard Briscoe[5], and Paolo Dini[6]

[1] Salzburg University of Applied Sciences & Technologies, A-5020 Salzburg, Austria
{thomas.heistracher, thomas.kurz, claudius.masuch}@fh-sbg.ac.at
[2] Soluta.net, 31037 Loria, Italy, pferronato@soluta.net
[3] Sun Microsystems Iberica, S.A, 08017 Barcelona, Spain, miguel.vidal@sun.com
[4] ISUFI, University of Lecce, 73100 Lecce, Italy, angelo.corallo@isufi.unile.it
[5] Imperial College London, SW7 2BT London, UK, gerard.briscoe@ic.ac.uk
[6] London School of Economics, WC2A 2AE London, UK, p.dini@lse.ac.uk

**Abstract.** In this paper we present ideas and architectural principles upon which we are basing the development of a distributed, open-source infrastructure that, in turn, will support the expression of business models, the dynamic composition of software services, and the optimisation of service chains through automatic self-organising and evolutionary algorithms derived from biology. The target users are small and medium-sized enterprises (SMEs). We call the collection of the infrastructure, the software services, and the SMEs a Digital Business Ecosystem (DBE).

## Introduction

From initially supporting human interactions with textual data and graphics, the Internet evolved to a global business platform; thus Web technologies have achieved a high utility for industry and businesses. Current web-based services connect different systems via information meta-models that facilitate the integration of the communication and execution layers. With the acquisition of these new principles and techniques, the Internet has been transformed from an information-only system into an integrated service platform. Thus the challenges for IT professionals are being shifted towards new levels. The possibilities made available by Web technologies displace the focus from implementing algorithms onto transcribing business needs. As a consequence, we can recognise the emergence of two different trends in software engineering.

First, the focus is still on the technical side: complex middleware systems require profound engineering knowledge, and enterprise application integration requires sound knowledge of different system architectures and technologies. For example, Web Services, as a low-level interfacing technology, provide basic integration capability for different execution environments like J2EE, .NET and so forth. Therefore, it is possible to integrate existing legacy systems step-by-step into existing business models and new architectural concepts. However, specifying only the technical functionality is not sufficient to describe a component or service as a whole. Thus, as a second trend, more abstract levels, i.e. illustrations of business processes, contract information, aspect functionality and so forth, have to be added to obtain and promote an overall understanding of the utility of services. This can enable the concatenation of single services into service chains and adaptive interaction and optimisation of the services within a distributed environment.

To meet both technical as well as business needs, more semantically rich abstract levels of description have to be defined in which the data and the code travel together from one extreme of the network to the other. Existing technologies like Web Services or ebXML are being used and extended already through enhancing technological dependencies and models by adding business- and domain-specific information. Therefore, the former specifications of technical needs are formulated via business modelling languages to describe the functionality of the software on a more abstract level. Although several initiatives can be observed now [1][2][3][4][5][6], the adaptive characteristics of these systems is still not a central concern of the current development processes. Coordination and adaptation of services as well as choreography descriptions for service chains are regarded as the most challenging fields of research in computer science and business over the next few years. In our work we complement the technologies and approaches mentioned above by taking inspiration from biological systems.

## Current Limitations and Challenges

This paper discusses on-going work in the Digital Business Ecosystem (DBE) Integrated Project under the 6$^{th}$ Framework Programme of Research in the Information Society Technologies thematic priority of the European Commission. The project aims to develop an open-source distributed environment that can support the spontaneous evolution and composition of (not necessarily open-source) software services, components, and applications. A central concern of the project is to provide a Business Modelling Language (BML) [7] that can usefully represent a wide spectrum of business transactions. We are therefore working with standard taxonomies for e-business model definitions [8] [9]. From these different approaches emerge criteria for identifying an e-business model based on how the relations between its actors are different combinations of coordination, cooperation and competition. The DBE approach wants to generate a general solution that, starting from a vision that considers SMEs as part of a specific network of business relations, defines an enabling infrastructure and a meta-service model. These could be adapted to sustain different e-business models also for the same firm, meaning that, from an e-business model point of view, the DBE approach allows SMEs to integrate their processes into different e-business networks and chains, reconfiguring their interfaces to interoperate within the same technological framework.

The flexibility afforded by the meta-modelling approach, coupled with the DBE infrastructural services, allows the optimisation of supply chains based on a much larger pool of potential players (i.e. other SMEs) from other sectors or geographical regions. We can describe this as a global optimisation that can bring very significant benefits to small companies, in contrast to the traditional approach of business management that benefits a local value chain at the expense of the single SMEs. The global view gives a sustainable improvement for all the business partners in the supply chain. Global optimisation of supply chains is influenced to a greater extent than local optimisation by dynamic effects. Such B2B interactions have so far been modelled only from a static point of view, or as a succession of pictures, or states, in which the dynamics of such complex systems is lost.

A new vision based upon the dynamics of biological systems could improve global optimisation mechanisms. As an extra benefit, the robustness of biological processes could be translated into more dependable and reliable business processes, benefitting especially SMEs. At a larger scale, the robustness of whole ecosystems could improve the stability characteristics of economies and of financial markets. The cost-effective nature of the DBE will allow SMEs to participate in this potentially planet-wide democratic business ecosystem, where each participant will have the same visibility in the network as the giant competitors.

In addition to greater flexibility and robustness of business networks and value chains, biological processes can guide the development of evolutionary algorithms for the incremental improvement of the business models and software service specifications through run-time feedback. This represents a bottom-up flow of information that complements the top-down BML approach mentioned above. These two opposite flows of information for representing software services are reconciled in the Service Manifest, as described in the next section.

## Architectural Vision

The DBE is not a regular project, it is a meta-project, in the sense that it is used to model "software projects" and their interactions. The functional and the technical specifications of such projects are unknown at design time in the DBE. The DBE is hence an environment where encapsulation, layering and meta-modelling are the prime building principles. Starting from this assertion, a common duality can be identified:
- Service Factory Environment
- Execution Environment

The *Service Factory Environment* in Fig. 1 is devoted to service definition and development. Clients of the DBE will use this environment to describe themselves and to generate software artifacts for subsequent implementation, integration and use. While in regular projects this phase is usually realised by relying on third-party tools, in the DBE we have to create our own tools. Moreover this is usually a single-step process, it terminates with the deployment of the application (except for a fraction of the effort for maintenance purposes), i.e. the end of the project. Here, on the other hand, the "Service Factory" is supposed to be never-ending. This parallel world is sometimes referred to as the "design-time of the DBE". The Service Factory Environment can be further divided into three other parts:

1. *Business specification*: where the business model of the service is realized (BML);
2. *Interface specification*: where the service is technically profiled and where an extra-functional definition is given (SDL);
3. *Coding*: where service code functionality is delegated to a legacy system or is implemented in its entirety (e.g. Java code, either generated or hand-written); all of the infrastructure code is auto-generated.

Fig. 1 shows the knowledge base that stores all the data types and models created as long-term memory. The short-term memory on the other hand stores the services when they are published and hence become available to the community.

The *DBE Service Execution Environment* in Fig. 1 is where services live. They are registered, deployed, searched, retrieved and consumed. This parallel world is sometimes referred to as the "run-time environment of the DBE". Even if from the technical point of view this environment could host the entire service implementation, it should be thought as a connector between the DBE and the SME internal applications: the DBE architecture is meant to be as non-intrusive as possible.

Fig. 2 shows a simplified picture of the run-time environment. The service is split in two parts: the "*smart proxy*" and the "*adapter*". The DBE application server (called Servent) on the left-hand side, once the proxy has been retrieved from the P2P network, wraps it and exposes it internally as a SOAP endpoint allowing it to be called by the legacy system. On the right-hand side the IT system of the supplier is shown to host the adapter that mediates the message calls to the legacy service provider system.

**Functional architecture**

The functional architecture can be broken down in the following categories:

- Structural services: to enable to DBE to work (ontology, P2P, modelling, security,...)
- Support services: to ease the development effort of partecipants (payment, Certification Authority, VOIP, email, information carriers)
- Basic services: integrated services (hotel reservation systems, booking, selling.. )
- Service chains: composed services

The respository (the knowledge base in Fig 1) is not structured. There will be no single reference "service model" or reference "meta-data repository" in the DBE to be maintained by some Commission. Enforcing a single meta-model has demonstrated in similar past projects to be a weak approach, leading to a single point of failure in the process of defining and maintaining services. The implementation was too complex and the expectations of its functionality too high: creating a unique reference data model satisfying the requirement of each service and vendor is nearly impossible, considering also the maintenance overhead.

Users will be encouraged to create their most adequate model by reuse or extension. An AI-based recommendation system will be able to match the user search criteria and to infer the correct types and interfaces based on the primary information and on the ontology. This resulting mechanism will allow the most suitable models to become a de facto standard through adoption by the user community and not because they are centrally imposed. In this distributed and de-centralised adaptive architecture, the BML framework will allow SMEs to interact and create business relations handling the parameters that define the firm, those that define services that allow to implement real-world SME services, and those that define agreement characteristics (i.e. volume discount policy, rules and parameters by which potential suppliers are admitted, payment methods, contract duration, warranty, import/export rules and limitations, etc).

In the DBE project the MDA approach will be extensively used for modelling. In fact, the meta-data repository will be MOF-compliant to enable easy data transformation, mapping and code generation. The BML framework is organised on three different layers: the level of meta-models (M2 level according to MDA), which lets DBE software users build BML models with the BML Editor; the level of BML models (M1), which enables the definition of service manifests through which general firm models are described; and the data level (M0), in which specific firms are described. The ontological M1 level is useful for associating a semantics to the specific SME descriptions, while the meta-ontological level M2 is useful for giving further semantics to BML and to grant coherence in order to avoid divergence in meanings of different domain ontologies.

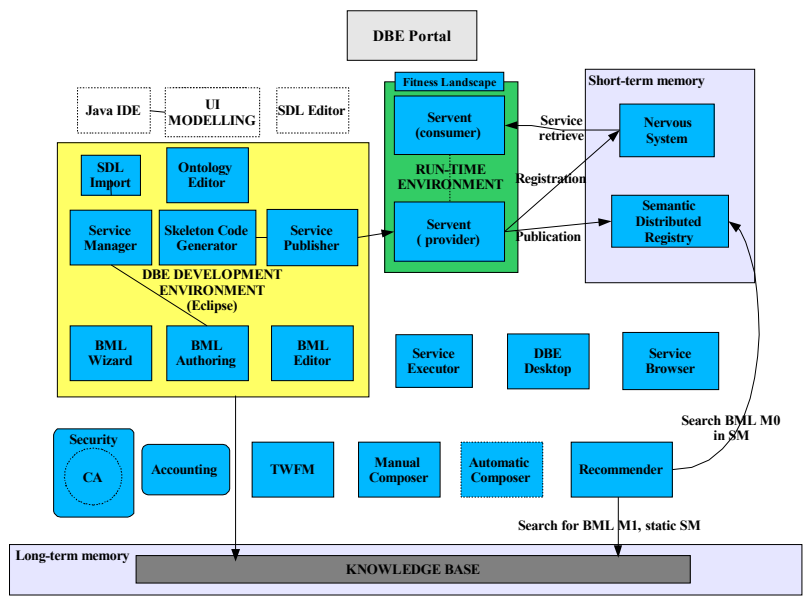

**Fig. 1.** Functional context diagram

The knowledge representing complex interactions in the SME business world abstracts from specific domains and contains general information, such as the role an enterprise can play in a business transaction (i.e. supplier or consumer), and will provide a well-defined set of parameters characterising European SMEs. Such parameters and their relations will be expressed as primitive concepts and relations in a top-level business meta-ontology. At this meta-ontological level, this knowledge will create constraints and rules for new application domains.  A level below, knowledge about specific domains will be represented by business domain ontologies defining specific businesses to support and with the insight necessary to build the service layer.  These ontologies create for each domain a repository of knowledge in terms of business strategy, business models and business processes, but also in terms of characteristics of firms. The BML will be developed starting from these ontologies. The ontology repository will represent a namespace for BML, allowing companies (by using the BML Editor) to enter information useful for advertising their services and for obtaining the right services through interaction with other SMEs in the DBE system.  BML will allow adding the business specification (pricing policy, contract duration, etc…) at the service level.

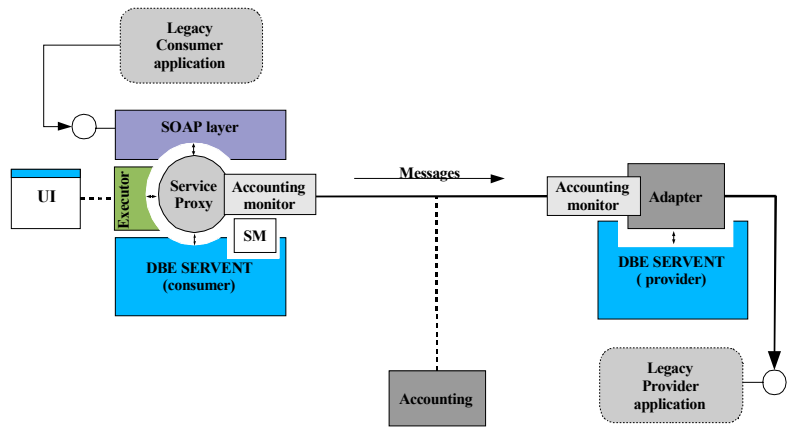

**Fig. 2.** Run-time overview of a Service proxy decomposed in its two main constituents

The complete definition of a service will be stored in a logical container called a Service Manifest, which contains the BML model and data, as well as the technical description needed for the run-time environment. The role played by the Service Manifest is equivalent to the role played by DNA in biological systems. Populations of Service Manifest chains have the capability to undergo mutation and replication. This coupled with a 'selection pressure' provides the precursors for evolution to occur, just as in nature with populations of organisms.  For example, fitness of the phenotype, which effects replication, is addressed by accepting run-time feedback to one of the infrastructural components of the DBE dubbed the "Population Evolver".

**Technical Viewpoint**

We tend to separate the overall technologies into two dimensions: supporting vs run-time. Although the project aims to reuse the run-time solutions also in the supporting dimension, there is not a complete overlap. For example the MDA approach is not enforced with SMEs, for whom it is completely transparent. On the other hand the Jini-based architecture is reused intensively in the DBE structural services like "accounting", "storage", and so forth.

The project commitment to the Open Source movement embraces the usage of components and tools released under free licenses. For example, as supporting technologies the project is going to take advantage of well-known available Open Source projects and standards such are Eclipse, Java, Jini, Apache/Tomcat, DSS, MDA-MOF-XMI/JMI, etc.

**Structural Viewpoint**

The architecture will support the distributed and interacting parts of the infrastructure and of the software components that inhabit it. Attention will be given to the crucial aspect of user interfaces to ensure that the DBE will indeed provide a competitive advantage to the participating SMEs while opening an innovative channel of software distribution for European software houses. The structural architecture will also explore different peer-to-peer network topologies for greater resilience, security, distributed storage to support the distributed intelligence and core functions and avoid a single point of failure. The architecture team is putting a lot of effort in the definition of a solution that will provide high availability to the supporting infrastructure of the DBE. The strategy is to enroll a P2P-based structure for distributing services and their business and technical descriptions. Any single failure in a part of the network or of a supporting node will not determine the failure of the entire ecosystem. The set of interactions is designed in a way that the system self-heals in case of partial failures and is resilient against perturbations, guaranteeing high availability. The DBE infrastructure will support the various development phases, such as bootstrap, production, deployment, test, etc, and will be populated with software services, SME software users and SME software providers.

As regards the P2P network topologies, our architectural approach is based on mapping businesses and interactions to the vertices and edges of a graph. When we add the "time" factor to static graph theory we have to take into account the dynamics of such graphs, as proposed by Barabási [10]. The DBE architecture will therefore address the business reality of changing relations between clients and suppliers, supporting dynamic supply chains composed of companies (vertices) and relations (edges) that change as a function of time. Each business should model its presence and activities as a small portion of the whole graph. For each element of the supply chain the graph (represented by means of BML) is composed of the relations between suppliers and clients. The graph should show a static and a dynamic view of the system.

In the static view the edges between vertices represent two kinds of relations: products/services flow and capital flow. The products/services flow will be represented by a collection of edges from the providers (purchase agreement) to the clients (sales agreement). The capital flow will be represented by a collection of edges from the clients (payable agreement) to the providers (payment agreement). The dynamic behaviour of this apparently random set of interactions will be rendered in the dynamic graph representation. This graph will represent transactions among business partners evolving until a final "architectural shape" (topology) emerges, in which a collection of business hubs will drive the main transactions, becoming the "de facto" core supporting companies of the whole system.

The set of companies who act as the core of the DBE will change over time as newcomers arrive, leading to better efficiency in the system. As an example, Google was not the first search engine to arrive on the Internet. At the time Google started, Yahoo and AltaVista were leaders of Web searching. When Google started to be linked to a huge number of pages in a more efficient way, its 'fitness' in the Internet improved and displaced other search engines

**Biological Concepts**

As stated, in the DBE project we are inspired by biology as the ultimate source of models for complex systems. The most intriguing aspect of biological systems that we would like to reproduce in software (at an appropriate level of granularity) is the construction of order. We recognise this class of phenomena as driven by physical interactions, symmetry laws of Nature, and global minimisation criteria (free energy, etc). However, in the DNA we find a symbolic level coexisting with and equally important to the physical level in the construction of order. Strictly speaking, symbols imply the assignment of meaning which can only happen through the (consensual) process of language formation—a social process according to Wittgenstein [11]. Thus, it may be more correct to speak of abstract structures or a dynamically inert code [12] that, through the entirely mechanical, probabilistic and "blind" (i.e. not conscious, sentient, or driven by a purpose) processes of gene expression and morphogenesis, come to "represent" phenotype (organism) structure and behaviour. What needs to be understood, therefore, is how a set of components can be "prepared" so that they will interact to form complex, dynamic, self-replicating structures in the presence of a steady supply of energy and atomic components.

If we can understand how all this might happen at a physical level we may be able to develop mathematical models that explain this "mechanism" and its robustness (meta-stability) properties. At the same time, we can't fail to notice that the same mechanism is mediated by recognisable patterns of components, on multiple scales. While we cannot go as far as claiming this dynamic hierarchy of recurring patterns to constitute a language, we can certainly ascribe it the status of information. The interesting possibility arises, therefore, to abstract a mapping between the physical processes that underlie the construction of order and a dynamic information management system. Our objective is to translate such an information management system into software algorithms and architectures, with the hope that the cooperative order-creating behaviour will be preserved.

As an initial step toward this ambitious vision, we can implement more elementary evolutionary algorithms to bring current technologies well beyond the UDDI-SOAP-WSDL interpretation of internet-based software services. For example, traditional software metrics as means of software quality assertion can be interpreted as analogous to estimation of biological fitness in software systems. Compared to biological fitness, however, software metrics are rather restricted fitness estimations. To express the dynamic interaction of the software services with the environment, additional and more sophisticated fitness properties that express and expose themselves at run-time (phenotype) must be discovered and taken into account. This approach renders necessary a significant extension of automated testing techniques, which in the DBE project will be reconciled with the biological viewpoint.

For a first estimation of fitness values within a distributed supply chain, the BML notation is suggested for the extension of current service descriptions. The biological simile here is a single cellular organism which couples with others to enable more complex processes that the individual alone cannot perform. Thus the users formulate their service needs via a BML request and the fitness of the service provided is determined by comparison of this BML request with the concatenated BML description of the service supply chain. A population of possible service supply chains will be evolved by selecting those individuals that the highest fitness to generate the optimal solution to the user request. For example, through an exchange of certain atomic services at run-time, applications can be improved or tuned to comply better to user requirements and changing needs.

There will be numerous SMEs, each making multiple requests to a large pool of atomic and aggregated services from which to evolve solutions. A distributed infrastructure, based on ecosystem theory to support the evolution of service supply chains for each SME, will support a 'habitat' for each SME. These 'habitats' provide a pool of software services and evolve a population of 'service supply chains' for each specific user request. The habitats will be interconnected with one another to create the software service ecosystem. The connections joining the habitats will be reinforced by successful software service utilisation and migration. This, along with similarities in user requests by different SMEs, will reinforce behavioural patterns and lead to clustering of habitats within the ecosystem, which can occur over industry sectors, geography, language, etc. This will form communities for more effective information sharing, the creation of niches and will improve the responsiveness of the system.

## Summary and Conclusions

This paper discusses how current technologies can be improved to extend their capabilities to fit into pervasive service architectures. We suggest a Digital Business Ecosystem which gives SMEs the possibility to concatenate their services within service chains, thereby improving their efficiency, by applying biological and physical concepts. There are several issues we have to address during our research activities in the coming years:

- Enable service specification, creation and composition via Business Process Descriptions using a Business Modelling Language
- Dynamic, automatic service composition by using self-organisation, evolution, and applying artificial intelligence
- Describe Software Components analogously to DNA
- Add replication rate dependence on usage feedback

- Create a software services ecosystem capable of supporting the evolution of populations of service supply chains: a system that is generic enough to be available to all SMEs, but also able to highly specialise and tailor solutions to the individuals SMEs.

Although some biological concepts have been mapped to software in the past, it is a novel approach to implement full evolutionary behaviour into software components, and use ecosystems theory for modelling the system architecture. In Nature, DNA (the genotype) encodes all the information necessary to develop the structure of the organism, and also controls the metabolism of the developed organism (the phenotype). One of the challenges we will address, therefore, will be to link current software services to their respective genotype and phenotype representations in order to unite the development and run-time environments into the same digital business ecosystem.

## Acknowledgements

This work is funded by the FP6 IST IP "Digital Business Ecosystem", contract number 507953.